\documentclass[aps,prd,superscriptaddress,twocolumn,showpacs,floatfix,linenumbers]{revtex4-1}

\usepackage{amsmath}
\usepackage{graphicx}
\usepackage{dcolumn}
\usepackage{bm}
\usepackage{array}
\usepackage{epstopdf}

\AppendGraphicsExtensions{.tif}

\newcolumntype{M}[1]{>{\centering\arraybackslash}m{#1}}
\newcolumntype{N}{@{}m{0pt}@{}}

\def\lsim{\mathrel{\raise.3ex\hbox{$<$\kern-.75em\lower1ex\hbox{$qf$}}}}
\def\gsim{\mathrel{\raise.3ex\hbox{$>$\kern-.75em\lower1ex\hbox{$\sim$}}}}

\begin{document}
\nolinenumbers

\title{Improved dark matter search results from PICO-2L Run 2}

\author{C.~Amole}
\thanks{Corresponding author}
\email{camole@owl.phy.queensu.ca}
\affiliation{Department of Physics, Queen's University, Kingston, K7L 3N6, Canada}
\author{M.~Ardid}
\affiliation{Departamento de Fisica Aplicada, Universitat Polit\`ecnica Val\`encia, Val\`encia, 46022, Spain}
\author{I.J.~Arnquist}
\affiliation{Pacific Northwest National Laboratory, Richland, Washington 99354, USA}
\author{D.M.~Asner}
\affiliation{Pacific Northwest National Laboratory, Richland, Washington 99354, USA}
\author{D.~Baxter}
\affiliation{Department of Physics and Astronomy, Northwestern University, Evanston, Illinois 60208, USA}
\author{E.~Behnke}
\affiliation{Department of Physics, Indiana University South Bend, South Bend, Indiana 46634, USA}
\author{P.~Bhattacharjee}
\affiliation{Saha Institute of Nuclear Physics, AstroParticle Physics and Cosmology Division, Kolkata, 700064, India}
\author{H.~Borsodi}
\affiliation{Department of Physics, Indiana University South Bend, South Bend, Indiana 46634, USA}
\author{M.~Bou-Cabo}
\affiliation{Departamento de Fisica Aplicada, Universitat Polit\`ecnica Val\`encia, Val\`encia, 46022, Spain}
\author{S.~J.~Brice}
\affiliation{Fermi National Accelerator Laboratory, Batavia, Illinois 60510, USA}
\author{D.~Broemmelsiek}
\affiliation{Fermi National Accelerator Laboratory, Batavia, Illinois 60510, USA}
\author{K.~Clark}
\affiliation{Department of Physics, University of Toronto, Toronto, M5S 1A7, Canada}
\author{J.~I.~Collar}
\affiliation{Enrico Fermi Institute, KICP and Department of Physics,
University of Chicago, Chicago, Illinois 60637, USA}
\author{P.~S.~Cooper}
\affiliation{Fermi National Accelerator Laboratory, Batavia, Illinois 60510, USA}
\author{M.~Crisler}
\affiliation{Fermi National Accelerator Laboratory, Batavia, Illinois 60510, USA}
\author{C.~E.~Dahl}
\affiliation{Department of Physics and Astronomy, Northwestern University, Evanston, Illinois 60208, USA}
\affiliation{Fermi National Accelerator Laboratory, Batavia, Illinois 60510, USA}
\author{M.~Das}
\affiliation{Saha Institute of Nuclear Physics, AstroParticle Physics and Cosmology Division, Kolkata, 700064, India}
\author{F.~Debris}
\affiliation{D\'epartement de Physique, Universit\'e de Montr\'eal, Montr\'eal, H3C 3J7, Canada}
\author{S.~Fallows}
\affiliation{Department of Physics, University of Alberta, Edmonton, T6G 2E1, Canada}
\author{J.~Farine}
\affiliation{Department of Physics, Laurentian University, Sudbury, P3E 2C6, Canada}
\author{I.~Felis}
\affiliation{Departamento de Fisica Aplicada, Universitat Polit\`ecnica Val\`encia, Val\`encia, 46022, Spain}
\author{R.~Filgas}
\affiliation{Institute of Experimental and Applied Physics, Czech Technical University in Prague, Prague, Cz-12800, Czech Republic}
\author{M.~Fines-Neuschild}
\affiliation{D\'epartement de Physique, Universit\'e de Montr\'eal, Montr\'eal, H3C 3J7, Canada}
\author{F.~Girard}
\affiliation{Department of Physics, Laurentian University, Sudbury, P3E 2C6, Canada}
\affiliation{D\'epartement de Physique, Universit\'e de Montr\'eal, Montr\'eal, H3C 3J7, Canada}
\author{G.~Giroux}
\affiliation{Department of Physics, Queen's University, Kingston, K7L 3N6, Canada}
\author{J.~Hall}
\affiliation{Pacific Northwest National Laboratory, Richland, Washington 99354, USA}
\author{O.~Harris}
\affiliation{Department of Physics, Indiana University South Bend, South Bend, Indiana 46634, USA}
\author{E.W.~Hoppe}
\affiliation{Pacific Northwest National Laboratory, Richland, Washington 99354, USA}
\author{C.~M.~Jackson}
\affiliation{D\'epartement de Physique, Universit\'e de Montr\'eal, Montr\'eal, H3C 3J7, Canada}
\author{M.~Jin}
\affiliation{Department of Physics and Astronomy, Northwestern University, Evanston, Illinois 60208, USA}
\author{C.~B.~Krauss}
\affiliation{Department of Physics, University of Alberta, Edmonton, T6G 2E1, Canada}
\author{M.~Lafreni\`ere}
\affiliation{D\'epartement de Physique, Universit\'e de Montr\'eal, Montr\'eal, H3C 3J7, Canada}
\author{M.~Laurin}
\affiliation{D\'epartement de Physique, Universit\'e de Montr\'eal, Montr\'eal, H3C 3J7, Canada}
\author{I.~Lawson}
\affiliation{Department of Physics, Laurentian University, Sudbury, P3E 2C6, Canada}
\affiliation{SNOLAB, Lively, Ontario, P3Y 1N2, Canada}
\author{A.~Leblanc}
\affiliation{Department of Physics, Laurentian University, Sudbury, P3E 2C6, Canada}
\author{I.~Levine}
\affiliation{Department of Physics, Indiana University South Bend, South Bend, Indiana 46634, USA}
\author{W.~H.~Lippincott}
\affiliation{Fermi National Accelerator Laboratory, Batavia, Illinois 60510, USA}
\author{E.~Mann}
\affiliation{Department of Physics, Indiana University South Bend, South Bend, Indiana 46634, USA}
\author{J.~P.~Martin}
\affiliation{D\'epartement de Physique, Universit\'e de Montr\'eal, Montr\'eal, H3C 3J7, Canada}
\author{D.~Maurya}
\affiliation{Bio-Inspired Materials and Devices Laboratory (BMDL), Center for Energy Harvesting Material and Systems (CEHMS), Virginia Tech, Blacksburg, Virginia 24061, USA}

\author{P.~Mitra}
\affiliation{Department of Physics, University of Alberta, Edmonton, T6G 2E1, Canada}
\author{S.~Olson}
\affiliation{Department of Physics, Queen's University, Kingston, K7L 3N6, Canada}
\author{R.~Neilson}
\affiliation{Department of Physics, Drexel University, Philadelphia, Pennsylvania 19104, USA}
\author{A.~J.~Noble}
\affiliation{Department of Physics, Queen's University, Kingston, K7L 3N6, Canada}
\author{A.~Plante}
\affiliation{D\'epartement de Physique, Universit\'e de Montr\'eal, Montr\'eal, H3C 3J7, Canada}
\author{R.B.~Podviianiuk}
\affiliation{Department of Physics, Laurentian University, Sudbury, P3E 2C6, Canada}
\author{S.~Priya}
\affiliation{Bio-Inspired Materials and Devices Laboratory (BMDL), Center for Energy Harvesting Material and Systems (CEHMS), Virginia Tech, Blacksburg, Virginia 24061, USA}
\author{A.~E.~Robinson}
\affiliation{Fermi National Accelerator Laboratory, Batavia, Illinois 60510, USA}
\author{M.~Ruschman}
\affiliation{Fermi National Accelerator Laboratory, Batavia, Illinois 60510, USA}
\author{O.~Scallon}
\affiliation{Department of Physics, Laurentian University, Sudbury, P3E 2C6, Canada}
\affiliation{D\'epartement de Physique, Universit\'e de Montr\'eal, Montr\'eal, H3C 3J7, Canada}

\author{A.~Sonnenschein}
\affiliation{Fermi National Accelerator Laboratory, Batavia, Illinois 60510, USA}
\author{N.~Starinski}
\affiliation{D\'epartement de Physique, Universit\'e de Montr\'eal, Montr\'eal, H3C 3J7, Canada}
\author{I.~\v{S}tekl}
\affiliation{Institute of Experimental and Applied Physics, Czech Technical University in Prague, Prague, Cz-12800, Czech Republic}
\author{E.~V\'azquez-J\'auregui}
\affiliation{Instituto de F\'isica, Universidad Nacional Aut\'onoma de M\'exico, M\'exico D. F. 01000, M\'exico}
\author{J.~Wells}
\affiliation{Department of Physics, Indiana University South Bend, South Bend, Indiana 46634, USA}
\author{U.~Wichoski}
\affiliation{Department of Physics, Laurentian University, Sudbury, P3E 2C6, Canada}
\author{V.~Zacek}
\affiliation{D\'epartement de Physique, Universit\'e de Montr\'eal, Montr\'eal, H3C 3J7, Canada}
\author{J.~Zhang}
\affiliation{Department of Physics and Astronomy, Northwestern University, Evanston, Illinois 60208, USA}

\collaboration{PICO Collaboration}
\noaffiliation

\date{\today}


\begin{abstract}
New data are reported from a second run of the 2-liter PICO-2L C$_3$F$_8$ bubble chamber with a total exposure of 129\,kg-days at a thermodynamic threshold energy of 3.3\,keV. These data show that measures taken to control particulate contamination in the superheated fluid resulted in the absence of the anomalous background events observed in the first run of this bubble chamber. One single nuclear-recoil event was observed in the data, consistent both with the predicted background rate from neutrons and with the observed rate of unambiguous multiple-bubble neutron scattering events. The chamber exhibits the same excellent electron-recoil and alpha decay rejection as was previously reported. These data provide the most stringent direct detection constraints on weakly interacting massive particle (WIMP)-proton spin-dependent scattering to date for WIMP masses $<$ 50\,GeV/c$^2$.

\end{abstract}

\maketitle


\section{Introduction}
The evidence for nonbaryonic dark matter is well established~\cite{PDG, dmevidence} and understanding the nature of particle dark matter is currently one of the most important quests in the field of particle physics~\cite{P5}. Weakly interacting massive particles (WIMPs) are a leading candidate for the cold dark matter in the universe and provide solutions for outstanding issues in both cosmology and particle physics~\cite{Jungman}. 

The sensitivity of a dark matter direct detection experiment depends on the WIMP mass and on the nature and strength of its coupling to atomic nuclei~\cite{wimptheory, wimpdetection, Snowmass}. Since theory provides little guidance as to the WIMP mass or coupling, it is important to explore multiple nuclear targets sensitive to various WIMP-nucleon couplings, including spin-dependent WIMP-proton, spin-dependent WIMP-neutron and spin-independent interactions. The $^{19}$F nucleus, because of its single unpaired proton and 100\,$\%$ isotopic abundance, provides a unique target to search for the spin-dependent WIMP-proton interactions. Experiments utilizing superheated fluorine-based liquids have consistently produced the strongest constraints on such interactions~\cite{PICO1, PICO60, PRD, previousPRL, PICASSOlimit, simple2014}.

The PICO Collaboration recently reported the observation of anomalous background events in dark matter search data with the 2-liter PICO-2L C$_3$F$_8$ bubble chamber~\cite{PICO1} deployed in the SNOLAB underground laboratory. The events were correlated in time with previous activity in the bubble chamber, and thus they were inconsistent with dark matter interactions and known backgrounds. Anomalous events with similar characteristics have also been reported in CF$_3$I bubble chambers~\cite{PICO60, PRD}. While analysis cuts based on the event timing were able to recover the dark matter sensitivity in Run-1~\cite{PICO1}, the presence of an unexplained background clearly indicated a limit to the technology and precluded scaling to a larger experiment.

PICO-2L Run-2 was initiated to explore the hypothesis that the anomalous background events observed in Ref.~\cite{PICO1} were caused by particulate contamination in the bubble chamber fluid. Particulate contamination is not present on the bubble chamber components following ultrasonic cleaning, yet it is expected from both the silica and stainless steel components of the bubble chamber. Stainless steel particulate is not produced in significant quantity during the assembly of the bubble chamber, but is expected to appear over the course of the run due to metal fatigue from the flexing action of the bellows and from corrosion. Silica particulate contamination is expected to arise primarily from fracturing of the mating surface of the silica inner vessel flange due to the mechanical stresses associated with its seal to the metal bellows flange. Stress fracturing~\cite{fracture} can result in significant production of silica particulate during the assembly of the vessel and, once initiated, stress corrosion fatigue is expected to provide an ongoing source of new silica particulate contamination. 

\section{Particulate Mitigation}
Measures taken to reduce the silica particulate contamination prior to Run-2 include the replacement of the quartz flange originally supplied on the fused silica inner vessel with a new flange fabricated from Corning 7980 Fused Silica~\cite{Corning}. In addition to being lower in radioactivity than quartz, the Corning material has fewer impurities, inclusions, and surface flaws and is therefore more likely to be resistant to stress fracturing~\cite{fracture,inclusions} and to the production of silica particles. A second measure was to modify the assembly sequence and fixtures to facilitate a more thorough rinse of the assembled vessel to remove silica particles that might have been generated during the assembly of the seal. Following the final rinse, the inner vessel assembly was dried using filtered gas flow and elevated temperature and it was evacuated and leak-checked using a turbo vacuum pump~\cite{turbo}, eliminating all exposure of the inner vessel to a scroll vacuum pump~\cite{Edwards} that was identified as a potential source of contamination in Ref.~\cite{PICO1}.

No measures were taken to mitigate the production of stainless steel particulate from the bellows prior to Run-2. Possible measures that were considered included specialized coatings to suppress particulate emission, a plastic bellows liner to contain the stainless steel particles, and replacement of the stainless steel bellows with a bellows formed from an alternative material. To avoid the possibility that the introduction of new construction materials might complicate the comparison of Run-2 to Ref.~\cite{PICO1}, the measures to mitigate the stainless steel contamination were deferred. For the same reason, a system developed for recirculation and filtering of chamber fluids was not implemented in Run-2. Consequently, the initial condition of the Run-2 bubble chamber was as identical as possible to the initial condition of Ref.~\cite{PICO1}, except for the reduction of silica and possible scroll pump particulate contamination, allowing for a direct comparison free from systematic differences.

Additional measures were also taken to reduce the agitation of the chamber to encourage settling of particulate, and to avoid stirring up any particles that might have settled out on the bubble chamber surfaces or the fluid interface. These measures include a careful optimization of triggering, expansion, and compression parameters, increasing the compression time between bubble nucleation events, and raising the pressure of the chamber from 31.1 psia, as in Ref.~\cite{PICO1}, to 37.2 psia, reducing the volatility of bubble growth. The Run-2 temperature was correspondingly increased in order to maintain the same 3.3\,keV thermodynamic energy threshold as Ref.~\cite{PICO1}.

\section{Other Modifications}

Several technical improvements unrelated to background reduction were implemented to improve the performance of the bubble chamber for Run-2. The number of temperature sensors was doubled and additional cooling was added to the top flange of the pressure vessel and to the camera enclosures to improve temperature uniformity across the active volume. Modifications were made to add over-voltage protection to the lead zirconate acoustic transducers and their number was increased from three to six to address a reliability problem encountered in Ref.~\cite{PICO1}. The VGA resolution cameras (491x656) used in~\cite{PICO1} were replaced with higher-resolution (1280x1024) devices to improve the spatial resolution of bubble position reconstruction.

\section{Operations}
The target mass of 2.91\,$\pm$\,0.01\,kg of C$_3$F$_8$ was kept in a superheated state at a temperature of 15.8\,$^{\circ}$C and a pressure of 37.2\,psia. For these run conditions, the thermodynamic threshold energy is estimated using the Seitz $``$hot spike" model~\cite{seitztheory} and is calculated to be 3.3\,$\pm$\,0.2(exp)\,$\pm$\,0.2(th)\,keV, with the experimental uncertainty originating from the uncertainty in temperature (0.3\,$^{\circ}$C) and pressure (0.7\,psi) and the theoretical uncertainty attributed to the thermodynamic properties of C$_3$F$_8$. The Run-2 thermodynamic threshold is equivalent to the lowest threshold reported in Ref.~\cite{PICO1} but at a higher temperature and pressure. The gross activity of the chamber in Run-2, measured by the number of expansions and the mean superheat time per expansion was comparable to Ref.~\cite{PICO1}. 

A total of 66.3 live-days of WIMP search data was collected at the 3.3\,keV thermodynamic threshold between June\,12 and September\,25, 2015. During this time, the detector was twice exposed to an AmBe calibration source to monitor the response to nuclear recoils from neutrons, and twice to a $^{133}$Ba source to evaluate the response to gamma-induced electron recoils. Data collected within 24 hours after any technical interruption were not included in the WIMP search. 

\section{Analysis}

The data analysis presented here uses techniques similar to those described in Ref.~\cite{PICO1}. All the neutron calibration data were scanned by eye to check the bubble multiplicities and the identified single-bubble events were used to evaluate the efficiency of the data analysis cuts. 

\begin{figure}[t]
\includegraphics[width=240 pt,trim=0 0 0 0,clip=true]{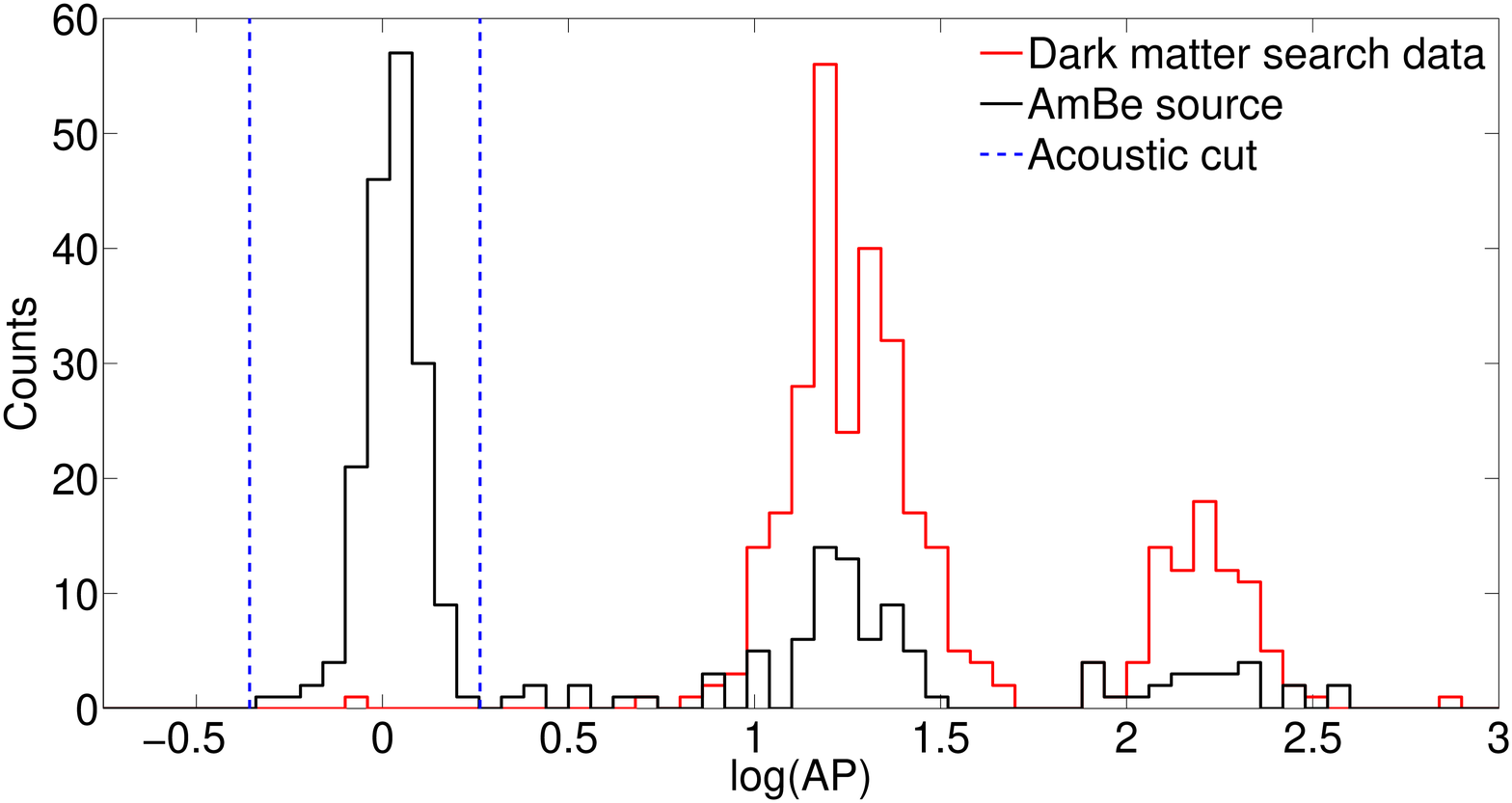}
\caption{\label{fig:AP} 
AP distributions (in log scale) of the single-bubble events originating within the optical fiducial volume for neutron calibration data (black) and WIMP search data (red). The signal region in AP for single nuclear recoils is indicated between the dashed blue lines. In both the calibration and WIMP search data, the two peaks at higher AP are from $^{222}$Rn chain alphas, with higher-energy alphas from $^{214}$Po decay producing larger acoustic signals~\cite{PICO1,PICO60}. The observed rate of alpha decays is consistent between WIMP search and neutron calibration data. 
}
\end{figure}

A set of data quality cuts was applied to remove events with failed optical reconstruction (bubble position and/or multiplicity), excessive acoustic noise, or poor agreement in the evaluated time of the bubble nucleation from the six acoustic transducers. The combined efficiency of the data quality cuts was 0.85\,$\pm$\,0.02. The acoustic analysis was performed using a procedure described in Ref.~\cite{PRD}, and the same acoustic parameter (AP) cut range of 0.7\,$<$\,AP\,$<$\,1.3 as in Ref.~\cite{PICO1, PRD, previousPRL} was adopted. The AP distributions for WIMP search and calibration data are shown in Fig.~\ref{fig:AP}. The AP cut has an acceptance of 0.94\,$\pm$\,0.02 for neutron-induced single-bubble events and an alpha rejection of $>$\,98.8$\%$ (90$\%$\,C.L.). An optical-based fiducial volume cut was derived such that less than 1$\%$ of the events originating at the interfaces (between C$_3$F$_8$, water buffer and glass walls) were accepted to be in the fiducial bulk volume and had an efficiency of 0.84\,$\pm$\,0.01.

The total acceptance for single-bubble nuclear-recoil events including data quality, AP, and fiducial cuts in this run was 0.67\,$\pm$\,0.03, resulting in a total exposure after cuts of 129\,kg-days. The position and acoustic resolution were significantly improved for Run-2, resulting in higher fiducial and AP cut efficiencies. However, the acceptance of the data quality cuts, and the total acceptance, was lower than in Ref.~\cite{PICO1} due to water droplets on the inside wall of the inner vessel compromising the optical reconstruction of a fraction of the events, and additional transient acoustic noise.

To search for neutron-induced multiple-bubble events in the WIMP search data, all events for which more than one bubble is reconstructed in one or both of the camera images were manually scanned. The acceptance of this selection criterion was determined using the neutron calibration data to be 0.93\,$\pm$\,0.01. This is substantially higher than the acceptance for single nuclear-recoil events since no acoustic or fiducial cuts are needed to identify multiple-bubble events.
 
\section{Backgrounds}
A constant rate (4\,cts/day) of AP-tagged alpha decay events was observed, similar to Ref.~\cite{PICO1}. Based on detailed Monte Carlo simulations, the background contribution from ($\alpha$,n) and spontaneous fission neutrons was predicted to be 0.008(0.010)\,counts/kg/day for single(multiple)-bubble events, with a total uncertainty of 50\%. This is higher than the estimate from Ref.~\cite{PICO1}, due to the addition to our simulation of ($\alpha$,n) reactions on $^{14}$N from radon-chain decays in air within the neutron shielding. The background model predicts 1.0(1.8) single(multiple)-bubble events from neutrons after all cuts. Fewer than 0.02 electron-recoil events were expected, based on a measurement of 4 candidate events during 12.2 live-days of exposure to a 1\,mCi $^{133}$Ba source coupled with a Monte Carlo simulation in GEANT4~\cite{GEANT4} of the natural gamma flux at the location of the chamber~\cite{gamma, alan}. The $^{133}$Ba calibration result corresponds to a measured efficiency of (2.2\,$\pm$\,1.2\,$)\times$10$^{-11}$ for electron recoils in C$_{3}$F$_{8}$ at a 3.3\,keV thermodynamic threshold.

\section{Results}
\begin{figure}
 \centering
\begin{tabular}{ll}
 \includegraphics[width=120 pt,trim=0 0 0 0,clip=true]{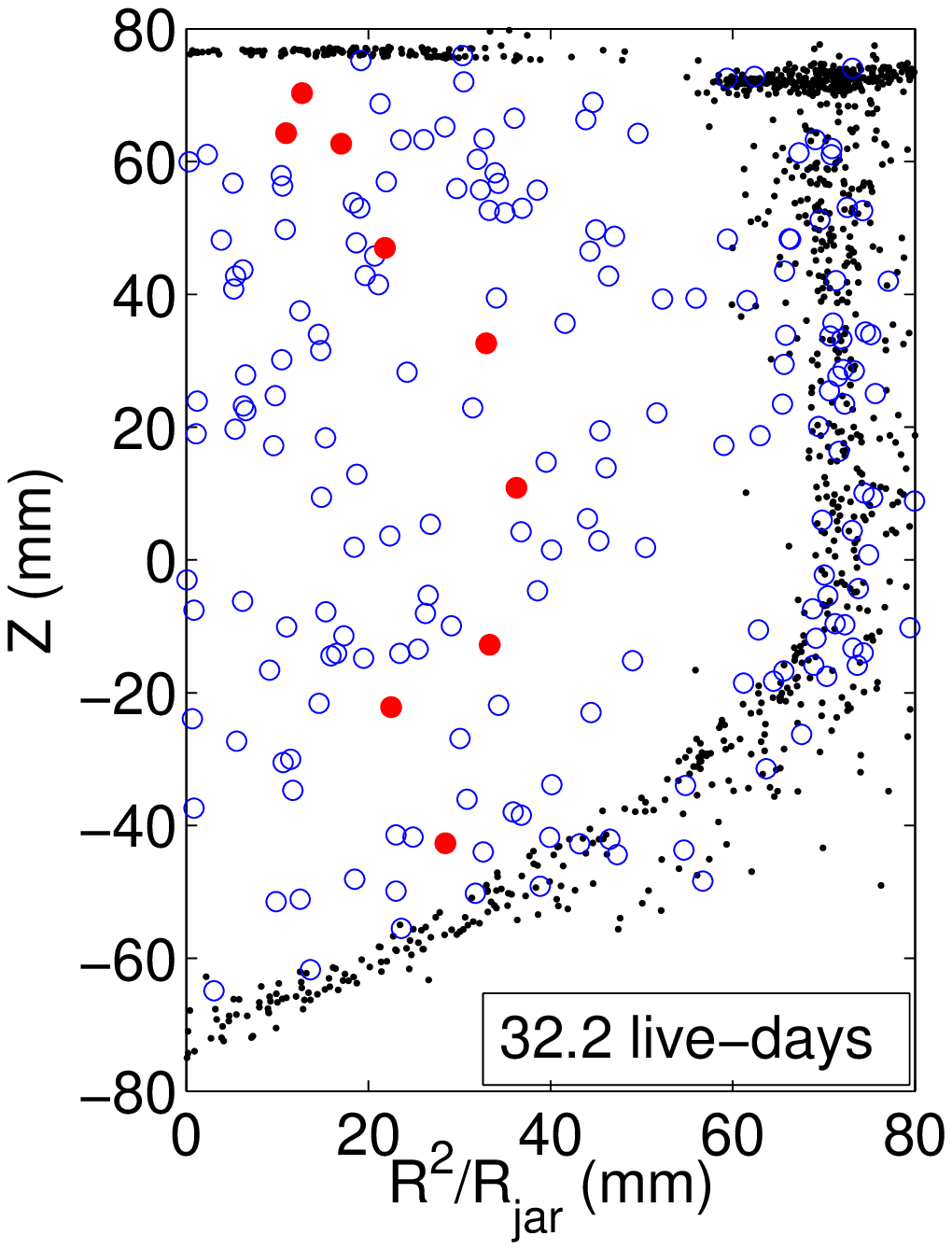}
 \includegraphics[width=120 pt,trim=0 0 0 0,clip=true]{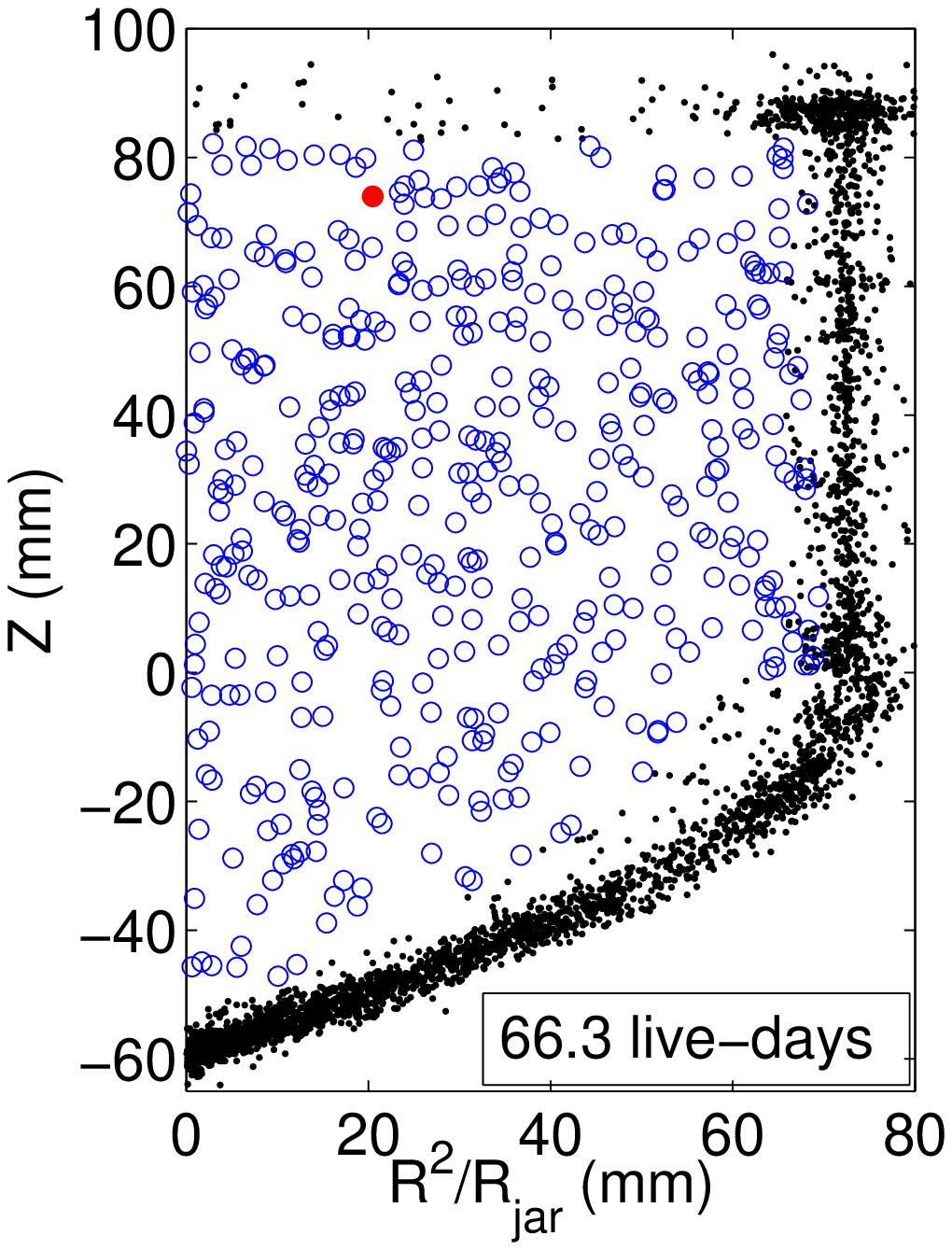}
\end{tabular}
\caption{\label{fig:xyz} Spatial distribution of bubble events in the 3.3\,keV WIMP search data for Run-1~\cite{PICO1} (left, 32.2 live-days) and Run-2 (right, 66.3 live-days). Z is the reconstructed vertical position of the bubble, R is the distance from the center axis and R$_{\textrm{jar}}$ the nominal inner radius of the silica jar (72.5\,mm). Red filled circles are WIMP-candidate events in the fiducial bulk volume, blue open circles are alpha-induced bulk events, and black dots are non-bulk events. The rate of pressure rise, measured by an AC-coupled transducer, was used for the fiducial volume cut in Ref.~\cite{PICO1}. An identical transducer installed for Run-2 failed during commissioning, and the Run-2 fiducial volume cut is entirely based on the improved optical reconstruction.}

\end{figure}

A total of 1(3) single(multiple)-bubble nuclear-recoil events were observed in the 129\,kg-day exposure. These data show the absence of the anomalous background events observed in the first run~\cite{PICO1} of PICO-2L (Fig.~\ref{fig:xyz}). The observed rate of both single- and multiple-bubble nuclear-recoil events is consistent with the expected background from neutrons. No neutron background subtraction is attempted, and the WIMP scattering cross-section upper limits reported here are simply calculated as the cross sections for which the probability of observing one or fewer signal events in the full 129 kg-day exposure is 10\,$\%$.

\begin{figure}
\includegraphics[width=240 pt,trim=0 0 0 0,clip=true]{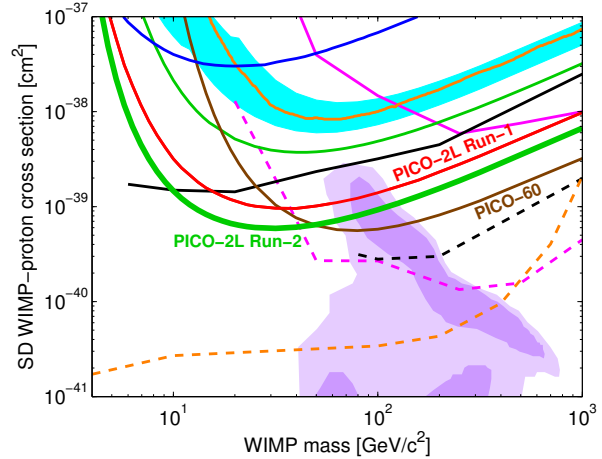}
\caption{\label{fig:SDplot} 
The 90\,$\%$\,C.L. limit on the SD WIMP-proton cross section from Run-2 (Run-1~\cite{PICO1}) of PICO-2L is plotted in green (red), along with limits from PICO-60 (brown), COUPP-4 (light blue region), PICASSO (dark blue), SIMPLE (thin green), XENON100 (orange), IceCube (dashed and solid pink), SuperK (dashed and solid black) and CMS (dashed orange)~\cite{PICO60,PRD,PICASSOlimit,simple2014,XENON100_SD,ICECUBElimit,SKlimit,SKlimit2,CMSmonojet}. For the IceCube and SuperK results, the dashed lines assume annihilation to $W$ pairs while the solid lines assume annihilation to $b$ quarks. Comparable limits assuming these and other annihilation channels are set by the ANTARES, Baikal and Baksan neutrino telescopes~\cite{Antares,Baksan,Baikal}. The CMS limit is from a monojet search and assumes an effective field theory, valid only for a heavy mediator~\cite{acc1,acc2}. Comparable limits are set by ATLAS~\cite{ATLASmonojet,ATLASheavyquark}. The purple region represents the parameter space of the constrained minimal supersymmetric standard model of Ref.~\cite{SDblob}.}
\end{figure}

\begin{figure}
\includegraphics[width=240 pt,trim=0 0 0 0,clip=true]{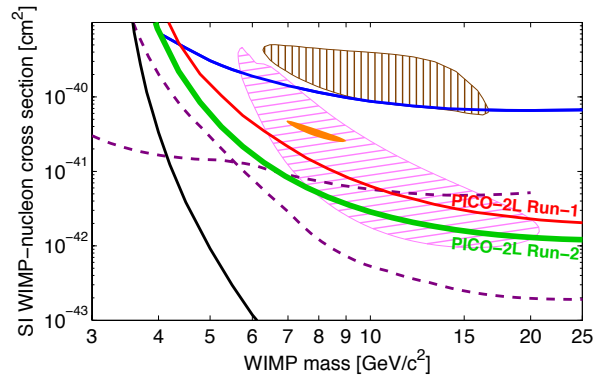}
\caption{\label{fig:SIplot} (Color Online)
The 90$\%$\,C.L. limit on the SI WIMP-proton cross-section from Run-2(Run-1~\cite{PICO1}) of PICO-2L is plotted in green(red), along with limits from PICASSO (blue), LUX (black), CDMSlite and SuperCDMS
(dashed purple)~\cite{PICASSOlimit,Lux2,CDMSlite2,SuperCDMS}. Similar limits that are not shown for clarity are set by XENON10, XENON100 and CRESST-II~\cite{XENON10, XENON100,CRESST2}. Allowed regions from DAMA (hashed brown), CoGeNT (solid orange), and CDMS-II Si (hashed pink) are also shown~\cite{DAMA, CoGeNT, CDMSSi}.
}
\end{figure}

The same conservative nucleation efficiency curves are used as in Ref.~\cite{PICO1}, with sensitivity to fluorine and carbon recoils above 5.5\,keV. The standard halo parametrization~\cite{lewinandsmith} is adopted, with $\rho_{\textrm{D}}$=0.3\,GeV$c^{-2}$cm$^{-3}$, v$_{\textrm{esc}}$ = 544\,km/s, v$_{\textrm{Earth}}$ = 232\,km/s, v$_{\textrm{o}}$ = 220\,km/s, and the spin-dependent parameters are taken from Ref.~\cite{spindependentcouplings}. Limits at the 90$\%$\,C.L. for the spin-dependent WIMP-proton and spin-independent WIMP-nucleon elastic scattering cross sections are calculated as a function of WIMP mass and are shown in Figs.~\ref{fig:SDplot} and~\ref{fig:SIplot}. These limits indicate an improved sensitivity to the dark matter signal compared to the previous PICO-2L run and are currently the world-leading constraints on spin-dependent WIMP-proton couplings for WIMP masses $<$\,50\,GeV/c$^2$. For WIMP masses higher than 50\,GeV/c$^2$, only the constraints from PICO-60~\cite{PICO60} are stronger.

\section{Discussion}
These data demonstrate the excellent performance of the PICO detector technology and provide strong evidence that particulate contamination suspended in the superheated fluid is the cause of the anomalous background events observed in the first run of this bubble chamber. Preliminary indications suggest that the radioactivity present in the particulate may be insufficient to account for the events as originating with alpha decays, so the bubble-nucleation mechanism associated with the particulate contamination is still unknown. Nonetheless, the identification of particulate contamination as the origin of the anomalous background events observed in Ref.~\cite{PICO1} provides the critical engineering guidance needed to develop a larger-scale background-free experiment.

\section{Acknowledgments}
The PICO Collaboration thanks SNOLAB for their exceptional laboratory space and technical support. We also thank Fermi National Accelerator Laboratory (Contract No. DE-AC02-07CH11359) and Pacific Northwest National Laboratory for their support. This work is supported by the National Sciences and Engineering Research Council of Canada (NSERC), the Canada Foundation for Innovation (CFI), the National Science Foundation (NSF) under the Grants PHY-1242637, PHY-0919526, PHY-1205987 and PHY-1506377, and by the U.S. Department of Energy under award DE-SC-0012161. We also acknowledge the support of Department of Atomic Energy (DAE), Government of India, under the Center of AstroParticle Physics II project (CAPP-II) at Saha Institute of Physics (SINP); the Czech Ministry of Education, Youth and Sports (Grant LM2015072); the Spanish Ministerio de Economia y Competitividad, Consolider MultiDark (Grant CSD2009-00064) and DGAPA-UNAM through grant PAPIIT No. IA100316.

\end{document}